\documentstyle[11pt,appb,epsf]{article}

\def\bq{\begin{equation}}
\def\eq{\end{equation}}
\def\bea{\begin{eqnarray}}
\def\eea{\end{eqnarray}}
\def\PhiL{\Phi_L}\def\PhiH{\Phi_H}\def\F{{\cal F}}

\newcommand{\eqref}[1]{(\ref{#1})}   

\def\roughly#1{\mathrel{\raise.3ex\hbox{$#1$\kern-.75em%
\lower1ex\hbox{$\sim$}}}}

\def\lsim{\roughly<}
\def\gsim{\roughly>}
\begin{document}
\title{EFFECTIVE FIELD THEORIES\\ FOR NUCLEI
AND DENSE MATTER\thanks{Invited talk at the Cracow Conference on
Structure of Mesons, Baryons and Nuclei, May 26-30, 1998, Cracow,
Poland} }
\author{Mannque Rho
\address{
Service de Physique Th\'eorique
\\CE Saclay
\\F-91191 Gif-sur-Yvette, France
\\ e-mail: rho@spht.sacaly.cea.fr
\\}
}
\maketitle
\begin{abstract}
A recent development on the working of effective field theories in
nuclei and in dense hadronic matter is discussed. We consider two
extreme regimes: One, dilute regime for which fluctuations are made
on top of the matter-free vacuum; two, dense systems for which
fluctuations are treated on top of the ``vacuum" defined at a given
density, with masses and coupling constants varying as function of
matter density (``Brown-Rho scaling"). Based on an intricate -- as
yet mostly conjectural -- connection between the in-medium
structure of chiral Lagrangian field theory which is a beautiful
effective theory of QCD and that of Landau Fermi liquid theory
which is an equally beautiful and highly successful effective
theory of many-body systems, it is suggested that a chiral
Lagrangian with Brown-Rho scaling in the mean field is equivalent
to Fermi-liquid fixed point theory. I make this connection using
electroweak and strong responses of nuclear matter up to nuclear
matter density and then extrapolating to higher densities
encountered in heavy-ion collisions and compact stars.
\end{abstract}

\section{Introduction}

Effective field theories (EFTs) are a powerful tool not only in
particle and condensed matter physics\cite{effective,pol} where
they are more extensively studied but also more recently, in
nuclear physics\cite{weinberg,pmr,vankolck,ksw,lm,cohen,NRZ} where
phenomenological approaches have traditionally been amply
successful, thus drawing less attention to field-theory approaches.
There are two superbly effective field theories that are quite
relevant to nuclear physics. One is chiral Lagrangian field theory
as a low-energy effective theory of QCD and the other is Landau
Fermi liquid theory as a semi-phenomenological theory for nuclear
matter. Both are beautiful examples of how effective field theory
works in hadronic systems. For nuclear many-body systems and most
of all for dense matter, both figure importantly. The first
involves what I would call ``chiral scale'' with the chiral cutoff
$\Lambda_\chi\sim 1$ GeV setting the scale below which the theory
is useful and the second involves ``Fermi-liquid scale'' set by
the Fermi momentum given by the density of the system.

In this talk, I would like to develop arguments that suggest that
combining the two effective theories leads naturally to the notion
of BR scaling \cite{BRscaling} which has recently found a simple
and striking application \cite{LKB} in the heavy-ion data of the CERES
collaboration \cite{ceres}. If the arguments are correct, the implication
is that what one usually attributes to change in the QCD vacuum
-- a quantity that is the focus of the present day nuclear and
hadronic physics
-- may be related, albeit indirectly, to many-body interactions on top of
the matter-free vacuum. This may be considered as a manifestation
of how two apparently different dynamical pictures represent the
same physical phenomenon or in the language of \cite{NRZ} a variant
of the Cheshire-Cat phenomenon. 
\section{Strategy for Effective Theory}

The idea of effective field theory is rather simple. Consider a
generic field $\Phi$ which we would like to study at an energy
scale less than a typical energy scale $\Lambda_1$. Let us divide
the field into the one we are interested in and the one we are not.
In terms of energy scales, the former corresponds to $\Phi_L$ for
$E<\Lambda_1$ and the latter to $\Phi_H$ for $E> \Lambda_1$,
$\Phi=\PhiL+\PhiH$. We are interested in the Feynman integral
$$Z=\int [d\Phi]
e^{iS[\Phi]}=\int [d\PhiL][d\PhiH]e^{iS[\PhiL,\PhiH]}.$$ Since we
are not interested in the degrees of freedom represented by
$\PhiH$, we will integrate it out of the Feynman integral. Define
\bea
e^{iS^{eff}[\PhiL]}=\int [d\PhiH]e^{iS[\PhiL,\PhiH]}\ ,
\eea
then the generating functional (when sources are suitably
incorporated) is $Z=\int [d\PhiL]e^{iS^{eff}[\PhiL]}$. This is an
exact result since we have not done anything other than redefine
things. Therefore we could have chosen the cutoff scale at
$\Lambda_2 <\Lambda_1$. In fact we could define the effective
action for any arbitrary scale by ``decimating" the cutoff. {\it If
everything is done correctly, physical quantities should not depend
upon how the $\Lambda_i$'s are chosen.} This statement is
translated into ``renormalization-group invariance." Now in our
case, although we know what the correct theory is (that is, QCD),
we do not yet know how to describe low-energy dynamics in terms of the
QCD variables (quarks and gluons). What we see in nature are
color-singlet hadrons. So the strategy is to write the effective
action at a given cutoff $\Lambda_i$ as an infinite series -- and
suitably truncate them -- in terms of known variables
\bea
S^{eff}_{\Lambda_i}=\sum_{n=0}^\infty C_n Q_n
\label{Seff}
\eea
where $Q$'s are local operators involving (observable) hadron fields written in
increasing power of momentum and/or of square of pion mass and
$C$'s are constants that are ``natural." In writing this expansion,
one appeals to symmetries such as Lorentz (or Galilei) invariance,
chiral invariance etc. In the usual chiral perturbation theory, the
expansion involves the pion and baryon fields with the power
$(\partial/\Lambda_\chi)^n$ and/or $(m_\pi^2/\Lambda^2_\chi)^n$.

How the effective action (\ref{Seff}) changes under ``decimation"
is expressed through Wilson's renormalization group-flow equation
\cite{pol}. This implies that the $\Lambda_i$-dependent
coefficients in (\ref{Seff}) satisfy the Wilson equation $
\frac{\partial C_i (\Lambda)}{\partial\Lambda}=\F_\Lambda (C_i)$
where $\F$ is a known function of $C_i$. In some cases, certain coefficients
stay constant under the decimation due to the presence of ``fixed
points." We shall see later that nuclear matter
is described by a fixed-point theory, with the nucleon  
effective mass and the four-Fermi quasiparticle 
interactions being fixed-point quantities.
\section{Two-Nucleon Systems}

I shall now illustrate how the above effective theory strategy
works in nuclear physics of two-body systems. All two-body systems
at very low energy are accurately known in nonrelativistic
phenomenological approach using two-body potentials. I propose that
they can provide a precision check of the theory that
we are developing.

Focusing on very low energy at an energy scale much less than the
pion mass, $m_\pi\approx 140$ MeV, we can integrate out all degrees
of freedom -- including pions -- other than the matter field,
namely, the nucleon field. Pions will be introduced later to go
higher order in the expansion. In the absence thereof, we can work
up to the next-to-leading order (NLO). We choose the cutoff
$\Lambda$ of the order of the pion mass. Define the four-point vertex
relevant to the process by
\begin{equation}
V({\bf q}) = \frac{4\pi}{M} \left(C_0 +
(C_2 \delta^{ij} + D_2 \sigma^{ij}) q^i q^j \right)+V_{EM},
\label{Vq}\end{equation}
where $V_{EM}$ is the electromagnetic interaction between two
protons which is of course known, $M$ is the nucleon mass and
$\sigma^{ij}$ is the rank-two tensor that is effective only in the
spin-triplet channel. The coefficients $C_{0,2}$ are (spin)
channel-dependent, and that $D_2$ is effective only in spin-triplet
channel. Thus there are five parameters; two in $^1S_0$ and three
in $^3S_1$ channel. In principle, these parameters are calculable
from a fundamental Lagrangian (i.e., QCD) but in practice,
nobody knows how to
do this. So in the spirit of EFTs, we shall fix them from
experiments. Since the explicit form of the regulator should not
matter\cite{lepage}, we shall choose the Gaussian form
$S_\Lambda({\bf p}) = \exp\left(- \frac{{\bf
p}^2}{2\Lambda^2}\right)$ where $\Lambda$ is the cutoff. As
mentioned, the cutoff is not a parameter to be fine-tuned; physical
quantities should not depend sensitively on it provided it is
correctly chosen for the scale involved.

Given the four-point function (\ref{Vq}), one can solve
Lippman-Schwinger equation or Schr\"odinger equation with
(\ref{Vq}) inserted as the kernel. This is strictly speaking not an
expansion in a rigorous accordance with the counting rule but one
can show that it is correct up to the order we are considering.
\begin{figure}[htb]
\begin{center}
\leavevmode
\epsfxsize=3in
\epsfysize=1.3in
\epsfverbosetrue
\epsffile{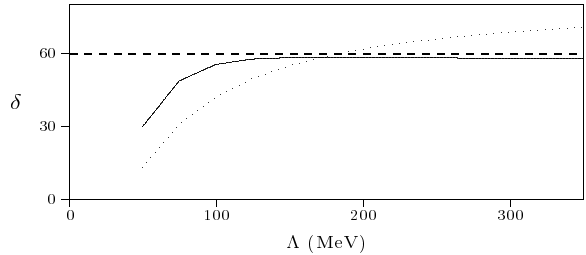}
\end{center}
\caption[deviate]{\protect \small
$np$ $^1 S_0$ phase shift (degrees) vs. the cutoff $\Lambda$ for a
fixed CM momentum $p= 68.5$ MeV. The NLO result is given by the
solid curve and the LO result by the dotted curve. The horizontal
dashed line is the result from the $v_{18}$ potential. Note the
$\Lambda$ independence for $\Lambda\gsim 150$ MeV.}\label{figure1}
\end{figure}

\begin{figure}[htb]
\begin{center}
\leavevmode
\epsfxsize=3in
\epsfysize=1.3in
\epsfverbosetrue
\epsffile{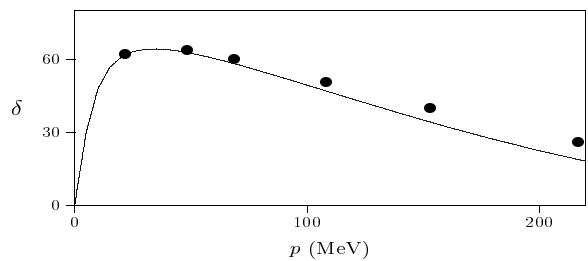}
\end{center}
\caption[phase]{\protect \small
$np$ $^1S_0$ phase shift (degrees) vs. the center-of-mass (CM)
momentum $p$. Our theory with $\Lambda=\Lambda_{Z=1}\simeq 172$ MeV
is given by the solid line, and the results from the Argonne
$v_{18}$ potential \cite{v18} (``experiments") by the solid dots.
(See \cite{pkmr1} for the precise definition of $\Lambda_{Z=1}$.)
As expected the theory starts deviating as the cutoff scale is
approached signaling that ``new physics" is setting in.
}\label{figure2}
\end{figure}

To see how the strategy works, let us consider low-energy
neutron-proton scattering. In Fig. \ref{figure1} is shown the $^1
S_0$ phase shift (in degrees) vs. cutoff for the scattering at a
fixed CM momentum of $p= 68.5$ MeV. One sees that below
$\Lambda\sim m_\pi$, the calculated phase shift varies rapidly and
disagrees with the experiment but once the cutoff is chosen at
about the pion mass, there is practically no cutoff dependence and
the theory agrees very well with the experiment as one increases
the cutoff. This therefore satisfies the condition for the
consistency of an effective theory. The second condition can be
seen in Fig. \ref{figure2}. For a given cutoff\footnote{See
\cite{pkmr1} for the precise procedure of picking this cutoff. One
should note that no fine-tuning is done here. The LO calculation
the cutoff $\Lambda_{Z=1}$ corresponds to the NLO calculation with
little dependence on cutoff in the sense of Fig. 1}, here taken at
$\Lambda=\Lambda_{Z=1}\simeq 170$ MeV, the theory agrees very well
up to $p\lsim 80$ MeV but beyond that it starts disagreeing. This
indicates that the theory breaks down as the momentum approaches
the cutoff. This may be due to the fact that higher order terms are
needed or new physics enters into the picture. This feature is
again required by the consistency of the effective theory.

This simple theory turns out to work extremely well for {\it all}
two-nucleon properties \cite{pkmr1}, namely, the properties of the
bound state deuteron, the radiative np capture
\bea
n+p\rightarrow d+\gamma\label{np}
\eea
and the solar proton fusion process
\bea
p+p\rightarrow d+e^+ +\nu_e.\label{pp}
\eea
\begin{table}
\caption[deuteronTable]{Deuteron properties and
the $M1$ transition amplitude entering into the $np$ capture for
various values of $\Lambda$.}\label{table1}
\begin{tabular}{cccccc}
$\Lambda$ (MeV)  & $198.8$ & $216.1$ & $250$ & Exp. & $v_{18}$\cite{v18} \\
\hline
$B_d$ (MeV)  & $2.114$ & $2.211$ & $2.389$ & $2.224$ & 2.224\\
$A_s$ ($\mbox{fm}^{-\frac12}$)
    & 0.877 & 0.878 & 0.878 & 0.8846(8) & 0.885 \\
$r_d$ (fm)
    & 1.960 & 1.963 & 1.969 & 1.966(7) & 1.967 \\
$Q_d$ ($\mbox{fm}^2$)
    & 0.277 & 0.288 & 0.305 & 0.286 & 0.270 \\
$P_D$ (\%)
    & 4.61 & 5.89 & 9.09 & $-$ & 5.76 \\
$\mu_d$
    & 0.854 & 0.846 & 0.828 & 0.8574 & 0.847 \\
$M_{\rm 1B}$ (fm)
    & 4.01 & 3.99 & 3.96 & $-$ & 3.98
\end{tabular}
\end{table}
As one can see in Table \ref{table1}, the NLO calculation gives a
remarkable agreement with all static properties of the deuteron,
again with little dependence on the cutoff. A much more striking
case is the radiative np capture process (\ref{np}) for which the
dominant contribution given by (\ref{Vq}) with $V_{EM}$ turned off
is found to agree precisely with the result of the Argonne $v_{18}$
potential \cite{v18}. The $\sim 10 \%$ exchange current
contributions that come at the next-to-next-to-leading
order (NNLO) can also be accurately calculated
\cite{pmr,pkmr1}. Taking into account inherent uncertainty in
short-distance physics which makes the main uncertainty in this process (in
nuclear physics language, this has to do with what is called
short-range correlation in the wavefunction), the calculated value
for the cross-section $\sigma_{ChPT}=334\pm 3$ mb is in perfect
agreement with the experimental value $\sigma_{exp}=334.2\pm 0.5$
mb. One could take this result as a ``first-principle" calculation.
This I believe is the first such calculation in nuclear physics.

The proton fusion process (\ref{pp}) plays a pivotal role for the
stellar evolution of main-sequence stars of mass equal to or less
than that of the Sun. The main contribution to the process comes
from (\ref{Vq}) (with the EM potential included) accounting for terms
up to NLO. Again
exchange currents enter at NNLO which can be incorporated in the
same way as in the np case, although the accuracy with which the
NNLO terms can be calculated is not as good as in the np case.
There are up to date no laboratory experimental data to check this prediction.
The inverse process to (\ref{pp}) is however presently being
measured and results will be forthcoming shortly. The only data so
far available come from helioseismology in the Sun \cite{helio}
which constrains the cross-section $S$ factor to
\bea
3.25 \lsim \frac{S(0)}{10^{-25}{\rm MeV-b}}\lsim 4.59.\label{helio}
\eea
The recent chiral perturbation calculation to NNLO \cite{pkmr2} --
which is an exact parallel to the np capture process --  gives
\bea
S(0)_{ChPT}=4.05 (1\pm 0.012)\times 10^{-25}\ \ {\rm MeV-b}.
\eea
This is consistent with the helioseismology (\ref{helio}) and
agrees with the value used in the physics of solar neutrino by
Bahcall and collaborators \cite{bahcallRMP} using the Argonne
$v_{18}$ potential
\bea
S(0)_{Bahcall}=4.00 (1\pm 0.007^{+0.020}_{-0.011})\times 10^{-25}\ \
{\rm MeV-b}.
\eea
\section{Infinite Nuclear Matter}
\subsection{Landau Fermi-liquid fixed points}

Going to infinite matter bypassing all intermediate-mass nuclei, we
encounter a new scale given by the Fermi sea occupied by
nucleons. We are still far from deriving the Fermi sea from a
chiral Lagrangian, not to mention from QCD. So I shall assume that
nucleons form a Fermi sea and occupy up to Fermi momentum $k_F$.
Consider excitations above and below the Fermi surface. Take a
cutoff for such excitations at say $\tilde{\Lambda}_1/2$ below and
above the Fermi sea and integrate out the excitations whose energy
is greater than $\tilde{\Lambda}_1$ and write effective actions
as described above. We may then proceed to do the ``decimation" as
above, but now around the Fermi surface. We shall call this
``Fermi-surface decimation." We learn from condensed matter systems
\cite{pol} where Fermi-liquid theory plays a prominent role that as
one scales down toward the Fermi surface, there are two families of
fixed points. Transcribed to nuclear matter, one of the two is the
nucleon effective mass $m_N^\star$ associated with the fixing of
the density of the system and the other is the four-Fermi
interaction that gives the Landau Fermi-liquid interaction ${\cal F}$.
That is to say, {\it nuclear matter can be described by Landau
Fermi-liquid fixed point theory}.
\subsection{Landau parameters and BR scaling}

It is possible to connect via BR scaling~\cite{BRscaling} the fixed
points of Landau Fermi liquid matter to the parameters of effective
chiral Lagrangians in dense medium. This can be done by looking at
the response of a nucleon on the Fermi surface to electroweak
fields \cite{FR96,frs98}.

By gauge invariance, the convection current of a nucleon on top of the Fermi
sea is given by the Landau-Migdal formula~\cite{migdal}
\bea
{\bf J}=g_l \frac{\bf p}{m_N}
\eea
where $g_l$ is
the orbital gyromagnetic ratio given by
\bea
g_l=\frac{1+\tau_3}{2}+\delta g_l
\eea
with $\delta g_l$ expressed in terms of Landau parameters $F_1$ and
$F_1^\prime$,
\bea
\delta g_l=\frac 16 (\tilde{F}_1^\prime
-\tilde{F}_1)\tau_3\label{deltagl1}
\eea
with $\tilde{F}=\frac{m_N}{m_N^\star} F$. On the other hand, chiral
and scale invariance of QCD implies~\cite{FR96,BRscaling}
\bea
\delta g_l=\frac 49 \left[\Phi^{-1} -1 -\frac 12 \tilde{F}_1^\pi\right]\tau_3
\label{deltagl}
\eea
where $\tilde{F}_1^\pi$ is the pionic contribution to the Landau
$F_1$ and $\Phi$ is the BR scaling parameter related to the ratio
of the quark condensate
$(\langle\bar{q}q\rangle^\star/\langle\bar{q}q\rangle_0)^n$ to some
power $n$, the dependence of which is model-dependent. $\Phi$ is
normalized such that at zero density it is equal to 1. Now the
Landau fixed-point mass
$\frac{m_N^\star}{m_N}=(1-\tilde{F}_1/3)^{-1}$ can also be
expressed in terms of the BR scaling and the pionic contribution,
$\frac{m_N^\star}{m_N}=(\Phi^{-1}-\tilde{F}_1^\pi/3)^{-1}$.
Comparing (\ref{deltagl1}) and (\ref{deltagl}) for $\delta g_l$, we
get
\bea
\tilde{F}_1-\tilde{F}_1^\pi\approx \tilde{F}_1^\omega =3 (1-1/\Phi)
\eea
where the superscript $\omega$ indicates contributions from {\it
all} massive isoscalar vector degrees of freedom, the most
important of which is the familiar $\omega$ meson. (All higher
energy mesons of the same quantum numbers are subsumed into that
factor.) In this simplified picture, the relevant long-wavelength
oscillation is given by the pion, $\tilde{F}_1^\pi$, and the
short-range by the $\omega$ meson, $\tilde{F}_1^\omega$.

From giant dipole excitations in heavy nuclei, we know that $\delta
g_l^p= 0,23 \pm 0.03$ for the proton~\cite{schumacher}. From this
we find that at normal density ($F_1^\pi$ is known by chiral
symmetry at any density)
\bea
\Phi(\rho_0)\approx 0.78.\label{Phirho0}
\eea
We will see later that this can be connected to the dropping vector
meson mass but for the moment we could simply relate it to the
ratio $f_\pi^\star/f_\pi$ and get the ratio from
Gell-Mann-Oakes-Renner mass formula applied to the mass of an
in-medium pion. Assuming that the effective pion mass increases a
bit in matter, one finds that the ratio at nuclear matter density
from the in-medium GMOR relation is $\sim 0.78$ and agrees with
(\ref{Phirho0}). This relation has been checked with axial-charge
transitions in heavy nuclei~\cite{warburton,KR,frs98}

An immediate check of (\ref{Phirho0}) is gotten by looking at the Landau mass
of the nucleon. For (\ref{Phirho0}), we get $m_N^\star (\rho_0)/m_N\simeq
0.70$. This agrees with the QCD sum-rule result~\cite{furnstahl}
$0.69^{+0.14}_{-0.07}$.

\subsection{Evidence from nuclear matter}

\def\del{\partial}
\def\la{\langle}\def\ra{\rangle}
The next relation we need to establish is between the scaling of
the meson masses and the BR scaling factor $\Phi$. To do this it
turns out to be most convenient to implement the scaling masses
into a chiral Lagrangian which in the mean field approximation
gives the nuclear matter ground state correctly. For this, write
the chiral Lagrangian truncated to the form of Walecka linear
$\sigma-\omega$ model (that is, drop all the fields that do not
enter in the mean field) as\footnote{The quantity $\rho$ that
figures in the parameters of the Lagrangian is not a number but an
operator whose mean field value is the matter density. How it is to
be treated is a bit subtle. Naive interpretation of the density
dependence of the mass leads to misleading results. See \cite{SMR}
for details.}
\bea
{\cal L}_{BR}& =&\bar{\psi}[\gamma_\mu (i\del^\mu-g_v^\star (\rho )
\omega^\mu )-M^\star (\rho )
+h\phi ]\psi\nonumber\\
&&+\frac12[(\del\phi )^2-m_s^{\star 2}(\rho )\phi^2]-\frac14 F_\omega^2
+\frac12 m_\omega^{\star 2}(\rho )\omega^2\label{model}
\eea
where $\psi$ is the nucleon field, $\omega_\mu$ the isoscalar
vector field, $\phi$ an isoscalar scalar field\footnote{Note that this 
scalar field is a
chiral singlet -- and not the fourth component of the chiral four-vector
of the linear sigma model -- 
to be consistent with chiral symmetry.} and the masses with
asterisk are taken to be BR-scaling. It has been
shown~\cite{SBMR,SMR} that this Lagrangian in the mean field
approximation gives {\it all} nuclear matter properties correctly
(including a low compression modulus in contrast to the linear
$\sigma-\omega$ Walecka model which differs from (\ref{model}) in that
the masses and coupling constants are non-scaling)
for the canonical values of free-space masses for the hadrons
provided the BR scaling
\bea
\Phi\approx
m_V^\star/m_V\approx M_N^\star/m_N\approx m_\sigma^\star/m_\sigma\approx
f_\pi^\star/f_\pi
\label{BR}
\eea
holds with $\Phi (\rho)\approx(1+0.28\rho/\rho_0)^{-1}$ and the
vector coupling scaling roughly the same way. As given, the scaling
of $\Phi$ is consistent with what we found in the baryon sector
(\ref{Phirho0}). Although the connection is somewhat indirect, it
is also possible to extract $\Phi$ from the QCD sum-rule
calculation of the $\rho$ meson in medium~\cite{hatsudalee,jin}. In
fact Jin et al find $m_\rho^\star (\rho_0)/m_\rho=0.78\pm 0.08$,
entirely consistent with (\ref{Phirho0}).

\subsection{Evidence from kaon-nuclear interactions}

There is yet another source for the scaling relation (\ref{BR})
that comes from the fluctuation of the BR scaling chiral Lagrangian
into the strangeness flavor direction. As discussed in
\cite{BR96,SBMR}, the BR scaling Lagrangian at tree order predicts
an attractive potential in the $K^{-}$-nuclear interaction which at
nuclear matter density comes to $\sim 190$ MeV. This attraction has
been seen in kaonic atom experiments. The recent analysis by
Friedman, Gal and Mares~\cite{FGM} gives the attraction of $185\pm
15$ MeV. This again supports the tree order calculation with BR
scaling fluctuating around the matter ground state. As discussed in
\cite{WRW}, the large attraction described in BR scaling can be
attributed to the higher chiral order effects that are not taken
into account in the conventional treatments.

\section{Dense Matter}
\subsection{Dileptons in heavy-ion collisions}

Fluctuating into non-strange directions, the effective Lagrangian
with BR scaling has been successfully applied to the dilepton data
of the CERES collaboration~\cite{ceres} by Li, Ko and
Brown~\cite{LKB}. The heavy-ion process involves densities
$\rho\sim 3\rho_0$, so a considerable extrapolation from nuclear
matter is required. In an extremely simplified form, the masses of
all hadrons drop linearly and become negligibly small at about
$3\rho_0$. The picture is then that near the chiral phase
transition the relevant degrees of freedom are the constituent
quarks, that is, weakly interacting quasiquarks. Since as argued
above, hadrons with BR scaling are quasiparticles at the density up
to about $\rho_0$, as density increases beyond $\rho_0$, the
effective degrees of freedom must crossover (possibly smoothly) in a
manner described by the NJL model from the hadron quasiparticles to
the quasiquarks forming the light-quark baryons and mesons up to
the chiral phase transition. This was the argument given in
\cite{BR96}. How this picture emerges in understanding the CERES
data will be discussed by Gerry Brown in the following talk.

\subsection{Kaon condensation in compact stars}

Fluctuated into the strangeness flavor direction, the dropping
$K^-$ mass discussed above leads in neutron star matter to
condensation of kaons at about $\sim 3\rho_0$ with important
consequences on the structure of compact stars~\cite{compact}.
Again the picture that emerges is that of the constituent quark.

\section{Conclusions}

In this talk, I argued that both dilute and dense hadronic systems
can be described in effective field theories. For the former,
the theory is defined in the matter-free vacuum and two-nucleon
systems, bound and elastic and inelastic scattering at low energy,
are accurately determined parameter-free when calculated up to NLO
in the chiral counting. For the latter, the ``decimation" at the
Fermi-sea scale is introduced and BR scaling is identified as a
means to {\it map} the mean-field chiral Lagrangian theory to Landau
Fermi-liquid fixed-point theory. The BR scaling for the nucleon is
checked with the electroweak responses of heavy nuclei and that for
mesons is checked with the fluctuations built on top of the
``vacuum" characterized by the density of the matter. The BR
scaling parameter $\Phi$ is shown to be related to the Landau interaction
parameter $F_1^\omega$ coming from massive isoscalar vector
degrees of freedom that underly {\it short-range interactions between 
nucleons}. This
implies that if the BR scaling is indeed connected to the vacuum
structure of QCD as argued here, {\it the change of the QCD vacuum
should be understandable in terms of interactions between hadrons,
at least up to a certain density below that of the chiral phase
transition}. This may be considered as a sort of Cheshire-Cat
phenomenon~\cite{NRZ}. It would be nice to quantify this statement.

Extrapolated into higher density regime in the most straightforward
way, the theory can be applied to dense matter in heavy-ion
collisions and in compact stars. As an effective theory, it is a
mean-field theory. Going beyond the mean field approximation and
calculating higher-order corrections remain to be formulated in a
systematic way.

Finally it is argued that as density is raised above normal matter
density, the correct degree of freedom should be the quasiquark and
hence there must be a change-over from hadronic Fermi liquid to
quark Fermi liquid of quasiquarks. Various phase transitions such
as the chiral or color superconductivity could be addressed from
the quark Fermi-liquid structure.

\vskip 0.4cm
\noindent
{\bf Acknowledgments}

It is a pleasure to dedicate this paper to Josef Speth on the
occasion of his 60th birthday. Josef and I had on various occasions --
and long before Landau-Migdal theory was widely recognized by the nuclear
physics community
-- exchanged our views on Fermi-liquid structure of nuclei and
nuclear matter and
the present paper is an unexpected and intriguing spin-off of the
ideas in a modern context. This paper is based on work done in
collaboration with Gerry Brown, Bengt Friman, Kuniharu Kubodera,
Dong-Pil Min, Tae-Sun Park and Chaejun Song whom I would like to
thank for discussions.

\def\pl{{ Phys. Lett.\ }}
\def\np{{ Nucl. Phys.\ }}
\def\pr{{ Phys. Rev.\ }}
\def\prl{{Phys. Rev. Lett.\ }}

\def\journal#1&#2(#3){{\sl #1} {\bf #2}(19#3)}
\def\etal{{\sl et al.}}
\def\UU{\relax}

\end{document}